\documentclass[useAMs,a4paper]{mn2e}
\usepackage{savesym}
\usepackage{graphicx}
\expandafter\let\csname equation*\endcsname\relax
  \expandafter\let\csname endequation*\endcsname\relax 
\usepackage{subfig}
\usepackage{amsmath}
\usepackage{amssymb}
\usepackage{verbatim}
\usepackage[yyyymmdd,hhmmss]{datetime}
\usepackage{array}
\usepackage{times}
\usepackage[total={17.8cm,24.0cm},centering]{geometry} 

\newcommand{\be}{\begin{equation}}
\newcommand{\beq}{\begin{equation}}
\newcommand{\ee}{\end{equation}}
\newcommand{\eeq}{\end{equation}}
\newcommand{\eea}{\end{eqnarray}}
\newcommand{\bea}{\begin{eqnarray}}

\newcommand{\dd}{\partial}

\newcommand\W {{W^r_{\ \phi}}}

\title[GR disc evolution ]{The general relativistic thin disc evolution equation}
\author [Steven A. Balbus]{Steven A. Balbus \thanks{E-mail:
steven.balbus@physics.ox.ac.uk} 
\\
Oxford Astrophysics. Denys Wilkinson Building, Keble Road, Oxford, OX1 3RH, United Kingdom}
\begin{document}

\date{}

\pagerange{\pageref{firstpage}--\pageref{lastpage}} \pubyear{2017}

\maketitle

\label{firstpage}

\begin{abstract} In the classical theory of thin disc accretion discs, the constraints of mass and angular momentum conservation lead to a  diffusion-like equation for the turbulent evolution of the surface density.  Here, we revisit this problem, extending the Newtonian analysis to the regime of Kerr geometry relevant to black holes.  A diffusion-like equation once again emerges, but now with a
singularity at the radius at which the effective angular momentum gradient passes through zero.   The equation may be analysed using a combination of WKB, local techniques, and matched asymptotic expansions.   It is shown that imposing the boundary condition of a vanishing stress
tensor (more precisely the radial-azimuthal component thereof) allows smooth stable modes to exist external to the angular momentum singularity, the innermost stable circular orbit, while
smoothly vanishing inside this location.   The extension of the disc diffusion equation to the domain of general relativity introduces a new tool for numerical and phenomenolgical studies of accretion discs, and may prove to be a useful technique for understanding black hole X-ray transients.   

\end{abstract}

\begin{keywords}
accretion, accretion discs --- black hole physics --- turbulence
\end{keywords}

\section{Introduction}

In thin disc accretion theory, the constraints of angular momentum and mass conservation may be combined into a single evolutionary equation for the disc surface density, a classic result first emphasised and discussed at length by  Lynden-Bell \& Pringle (1974) (see also the review of Pringle 1981).    In its original implementation, the equation takes the form of a diffusion equation, with a diffusion
coefficient proportional to an {\em ad hoc} turbulent viscosity.    Balbus \& Papaloizou (1999) later showed
how the same evolution equation emerges without the need to introduce an explicit viscosity.  By writing the velocity field as a sum of a mean plus a fluctuation (with vanishing mean), an effective diffusion coefficient emerges which is proportional to the correlation in the radial and azimuthal velocity fluctuations\footnote{The presence of a magnetic field
can be incorporated into this formalism, with the product of the radial and azimuthal Alfven velocities subtracted from the kinetic velocity fluctuations in the diffusion coefficient (resulting in an {additional} {\em positive} stress).}.   

The evolutionary equation has heretofore been used in the regime of Newtonian gravity (e.g. Pringle 1981). 
Solutions of the equation show that matter in accretion discs drifts inward, while angular momentum is transported outward, sustained by
a vanishingly small mass fraction of the disc.   The extension of the evolutionary equation to include general relativistic gravity has not yet been done, and it is not without interest.   It is the purpose of this paper to derive and analyse the general relativisitic version of the thin disc evolutionary equation.   We present a very general global asymptotic analysis (assuming small stress and/or rapid modal time scales), which can be equally well applied in the Newtonian limit.

The inner regions of neutron star and black hole discs are dynamically complex.  It would be naive to apply simple thin disc dynamics uncritically.    The formal thin disc problem is nevertheless quite interesting, first
as an illustration of how the diffusion dynamics breaks down at the innermost stable circular orbit (ISCO) of the disc, second of how
the diffusion equation extends to ISCO-free Kerr orbits in general relativity,  and third as a useful analytical tool for understanding numerical
simulations.     It is especially noteworthy that while the effective diffusion coefficient of the disc equation becomes singular at the ISCO, the solution is nevertheless mathematically well-behaved.   The global normal modes include exponentially growing modes confined to the zone
within the ISCO, which completely disrupt the interior disc structure, leaving the outer disc intact.  This is in accord with numerical simulations.   

The plan of this paper is as follows.   In \S 2 we first derive the form of the disc evolution equation that follows from the conservation
of particle number and the azimuthal component of the stress energy tensor.   This reduces to the Lynden-Bell---Pringle (1974) equation in 
the Newtonian limit.  A solution of the general equation is presented in \S3 for modes with exponential time dependence, using WKB, local analysis, and matched asymptotic expansions.    The cases of both finite and vanishing stress at the location of the ISCO are presented, and we argue that thin discs will evolve to a state at which the vanishing stress boundary condition is achieved.  We use the modal solutions to construct a general Green's function solution.    Finally, in \S4 we summarise the presentation.   This scope of this paper is to present a mathematical treatment of the equation.   Astrophysical applications will be explored
in a separate study.   

We observe the following conventions.   {\em The speed of light is set to unity throughout this work.}     Greek indices $\alpha, \beta, \gamma...$
generally denote spacetime coodinates.   The exception is $\phi$, which is always the azimuthal angular coordinate.  The time coordinate is labelled $0$.  The metric in local inertial coordinates is
$
g_{\alpha\beta} \rightarrow \eta_{\alpha\beta} = {\rm diag\ }(-1,1,1,1).
$
Other notation is standard: $G$ is the gravitational constant, $M$ the central black hole mass, and $r_g=GM$ the gravitational
radius.

\section{Fundamental equations}

\subsection{Conserved fluxes}

The two conserved quantities of interest are the particle number current $nU^\mu$, where $n$ is the rest 
frame number density and $U^\mu$ the contravariant 4-velocity, and the azimuthal component of the stress energy tensor $T^\mu_{\ \phi}$.   Following Page \& Thorne (1974), we will work in ``cylindrical Boyer-Lundquist'' $r, \phi, z$ coordinates in the Kerr metric, ultimately using the equations in their height-integrated form.   This involves ignoring
the higher order curvature terms of order $z^2/r^2$ near the equatorial plane.   We assume that the disc is axisymmetric and thin.  
The conservation equation of particle number before height integration is simply
\beq
(nU^\mu)_{;\mu} = 0,
\eeq
where the semi-colon denotes a covariant derivative.   If we now integrate over $z$,  and assume that the velocities are independent of height, the remaining 4-velocity components are
$U^0, U^r$, and $U^\phi$.  With rest mass per particle $m$, and $\Sigma$ the integrated column density
$$
\Sigma =m\int  n\, dz,
$$
the particle conservation equation for an axisymmetric disk is
\beq\label{partcon}
{1\over \sqrt{g}}\dd_\mu \left( \sqrt{g}\,\Sigma U^\mu\right) = U^0 \dd_t\Sigma +{1\over \sqrt{g}}\dd_r \left( \sqrt{g}\,\Sigma U^r\right) = 0
\eeq
where $g=|\rm{det}\,g_{\mu\nu}|$, the absolute value of the metric tensor determinant. 

The $\phi$ equation for conservation of the stress-energy tensor is $T^\mu_{\ \phi;\mu}=0$,
with $T_{\mu\nu}$ taking the form of an ideal fluid plus a contribution from the radiation field, denoted $\tau_{\mu\nu}$: 
\beq
T_{\mu\nu} =  g_{\mu\nu} P +(\rho+P) U_\mu U_\nu +\tau_{\mu\nu}.
\eeq
The radiation stress $\tau_{\mu\nu}$ is given by
\beq
\tau_{\mu\nu} = q_\mu U_\nu + q_\nu U_\mu,
\eeq
where $q_\mu$ is the radiative energy flow vector, which satisifies $q_\mu U^\mu = 0$ (Page \& Thorne 1974). 
We ignore (for the moment) the contribution of any additional stress tensor that may be present.   The angular momentum 
carried by radiated photons is not negligible when rotational velocities are of order the speed of light (Novikov \& Thorne 1973).   
Then, assuming axisymmetry $\dd_\phi = 0$, and defining
\beq
\sigma_{\mu\nu} = (\rho + P)U_\mu U_\nu + \tau_{\mu\nu},
\eeq
the conservation equation becomes
\beq\label{three}
0={1\over \sqrt{g}}\dd_\mu(\sqrt{g} \sigma^\mu_{\ \phi})-\Gamma^\lambda_{\mu\phi}\sigma^\mu_{\ \lambda}.
\eeq
Here $\rho$ is the rest energy density (including in principle a
thermal contribution, which is however ignored in the thin disk limit), $P$ is the thermal pressure (which shall likewise be ignored), and $\Gamma^\lambda_{\mu\phi}$ is the affine connection.   For axisymmetric $\dd_\phi=0$ metrics this is:
\beq\label{gam}
\Gamma^\lambda_{\mu\phi}= {1\over 2}g^{\lambda\alpha}( \dd_\mu g_{\alpha\phi}- \dd_\alpha g_{\mu\phi}).
\eeq
Therefore, for {\em any} symmetric tensor $\sigma^{\mu\nu}$, the combination $\Gamma^\lambda_{\nu\phi}\sigma_{\ \lambda}^\nu$ is
\beq\label{gammi}
{1\over 2}g^{\lambda\alpha}( \dd_\mu g_{\alpha\phi}- \dd_\alpha g_{\mu\phi})\sigma_{\ \lambda}^\mu=
{1\over 2}( \dd_\mu g_{\alpha\phi}- \dd_\alpha g_{\mu\phi})\sigma^{\mu\alpha} =0,
\eeq
since the metric derivatives are antisymmetric in $\alpha$ and $\mu$ while $\sigma^{\mu\alpha}$ is symmetric\footnote{Knowledgeable readers will recognise in equation (\ref{gammi}) a Killing vector calculation.  I thank C.\ Gammie for drawing my attention to this point.}.
By contrast, $g_{\mu\nu}$ is {\em not} independent of $r$, so that
\beq\label{req}
\Gamma^{\lambda}_{\mu r}\sigma^{\mu}_{\ \lambda} = {1\over 2}\sigma^{\alpha\mu}\dd_rg_{\alpha\mu}, 
\eeq
a result we use below.   Equation (\ref{three}) now reduces to:
\beq\label{am0}
{1\over \sqrt{g}}\dd_\mu(\sqrt{g} \sigma^\mu_{\ \phi})=0.
\eeq 


The disc turbulence is represented by writing the 4-velocity $U^\mu$ as
a mean flow $\bar U^\mu$ plus a fluctuation $\delta U^\mu$ with vanishing mean, $\overline{\delta U^\mu}=0$.
In particular, 
\beq
\overline{U^r U_\phi} = \bar U^r \bar U_\phi + \overline{ \delta U^r\, \delta U_\phi} \equiv  \bar U^r \bar U_\phi +\W
\eeq
The asymptotic scalings of the fluctuations satisfy:
\beq
\delta U_\phi \ll \bar U_\phi, \quad \bar U^r \ll \delta U^r \sim \delta U_\phi/r \ll  r \bar U^\phi,
\eeq
i.e., the orbital velocity and angular momentum are much larger than their associated fluctuations, and the inward mean radial drift velocity is yet an asymptotic order smaller than the fluctuations in either the radial velocity or orbital velocity.  The two $\delta U$ fluctuations are assumed to be of comparable order,
suitably dimensionalised.   In common with Newtonian theory (Balbus \& Papaloizou 1999), we expect $\bar U^r \bar U_\phi$ (the product of a zeroth order rotational velocity and a second order radial drift) to be of the same asymptotic order as  $\W$ (the product of two first order fluctuations). 
As always, it is important to distinguish contravariant $U^\phi$ (angular 4-velocity) from covariant $U_\phi$ (angular 4-momentum):
$$
U_\phi = g_{\phi 0} U^0 + g_{\phi\phi}U^\phi = g_{\phi 0} {dt\over d\tau}+ g_{\phi\phi}{d\phi\over d\tau},
$$
where we have ignored $U^r$ as negligibly small.

\subsection{``Stress by strain'' and radiation}
\subsubsection {Equilibrium models}

For
the equilibrium models under consideration, Page \& Thorne (1974) present a relationship between the disc shear, a tensor coupling like viscosity, and the energy radiated from its surface.   In our notation, this relation reads:
\beq\label{10b}
-\Sigma \W\bar U^0{d\Omega\over dr} =  2{\cal F},
\eeq
where 
\beq
\Omega = {d\phi\over dt}={d\phi\over d\tau}{d\tau\over dt}={\bar U^\phi\over \bar U^0}
\eeq
is the angular velocity measured by an observer at infinity,
and ${\cal F}$ is the radiated energy flux in the local rest frame.    In essence,
this states that the energy extracted from differential rotation and put into turbulent fluctuations is locally radiated away at the same rate.   
We will make use of this relation in \S 2.3 below, which also holds in our case because of the assumption that the thermal timescale is more
rapid than the evolutionary timescale.   
It is of some technical interest to revisit
this important relationship in more detail in an out-of-equilibrium context, which we do in the following section (see also Balbus \& Hawley 1998, Balbus \& Papaloizou 1999).   The reader willing to adopt equation (\ref{10b}) directly may wish to skip directly to \S 2.3 below, without loss of continuity. 

\subsubsection{Free energy from shear}

The radial $T^\mu_{\ r}$ conservation equation is given, with the help of equation (\ref{req}) and particle number conservation, by
\beq\label{10}
{\delta U^r\over \sqrt{g}}\dd_\mu(\sqrt{g}\rho U^\mu U_r+\sqrt{g}\tau^\mu_{\ r}) -{\rho \delta U^r\over 2} U^\alpha U^\mu\dd_r g_{\alpha \mu} = 0.
\eeq
We have multiplied by $\delta U^r$ with the aim of assembling a fluctuation energy equation; the radial component  $\tau^\mu_{\ r}$ of the radiation stress is small, but retained here to maintain a covariant
formulation.

Next, we write the $U$ velocities as a mean $\bar U$ plus fluctuating $\delta U$.    The largest contributions from the final term
of equation (\ref{10}) comprise the equilibrium
solution and are not of interest; they cancel out.   Retaining the next largest group of terms, our equation becomes
\beq\label{12}
{\delta U^r\over \sqrt{g}}\dd_\mu[\sqrt{g}\rho U^\mu (\bar U_r+\delta U_r)+\sqrt{g}\tau^\mu_{\ r}] -{\rho \delta U^r} \delta U^\alpha \bar U^\mu\dd_r g_{\alpha \mu} = 0,
\eeq
where, in the final term, we have used $g_{\alpha\mu}=g_{\mu\alpha}$ symmetry.   It is convenient for now to retain $U^\mu$ in the $\dd_\mu$ divergence term without separating its mean and fluctuation.  (The radiation stress $\tau^\mu_{\ \nu}$ will likewise contain a fluctuating $\delta U$ component, which is not shown explicitly.)  The term involving $\bar U_r$ is an asymptotic order
smaller than the others, and may be dropped.   We arrive at:
\beq
{\delta U^r\over \sqrt{g}}\dd_\mu(\sqrt{g}\rho U^\mu \delta U_r+\sqrt{g}\tau^\mu_{\ r}) -{\rho \delta U^r} \delta U^\alpha \bar U^\mu\dd_r g_{\alpha \mu} = 0,
\eeq

Next, the equation for $T^\mu_{\ \phi}$ (angular momentum conservation) is
\beq\label{13}
 {\delta U^\phi\over \sqrt{g}}\dd_\mu[\sqrt{g}\rho U^\mu (\bar U_\phi +\delta U_\phi)+\sqrt{g}\tau^\mu_{\ \phi}] = 0.
\eeq
Using particle number conservation and remembering that $\bar U_\phi$ depends only upon $r$, this becomes 
\beq\label{133}
 {\delta U^\phi\over \sqrt{g}}\dd_\mu[\sqrt{g}(\rho U^\mu\delta U_\phi+\tau^\mu_{\ \phi})] +\rho \delta U^\phi  \delta U^r\dd_r\bar U_\phi =0.
\eeq
Following the thin disc Novikov and Thorne (1973) models, the radiation flux $\tau^\mu_\phi$ is assumed to be dominated by its vertical $\tau^z_{\ \phi}$ component, and in particular by the
$q^z U_\phi$ term.    
We rewrite the last term to obtain: 
\beq\label{14}
 {\delta U^\phi\over \sqrt{g}}\dd_\mu[\sqrt{g}(\rho U^\mu\delta U_\phi+\tau^\mu_{\ \phi})] +\rho \delta U^\phi  \delta U^r\dd_r (g_{\phi\alpha}\bar U^\alpha) =0.
\eeq

The $T^\mu_{\ 0}$ equation is handled similarly:
\beq
\delta U^0{1\over \sqrt{g}}\dd_\mu\sqrt{g}[\rho U^\mu(\bar U_0+\delta U_0)+\tau^\mu_{\ 0}]=0, \ \ \ \ \ \ \ \ \ \ \ \ \ \ \ \ \ \ 
\eeq
which expands to 
$$
\delta U^0{1\over \sqrt{g}}\dd_\mu\sqrt{g}(\rho U^\mu\delta U_0+\tau^\mu_{\ 0}  ) +\rho\delta U^0\delta U^r\dd_r\bar U_0 =0,
$$
or
\beq\label{16}
\delta U^0{1\over \sqrt{g}}\dd_\mu\sqrt{g}(\rho U^\mu\delta U_0+\tau^\mu_{\ 0}) +\rho\delta U^0\delta U^r\dd_r(g_{\alpha 0}\bar U^\alpha) =0.
\eeq

The final $T^\mu_{\ z}$ equation is simple and straightforward, as there is by assumption no mean $z$ flow:
\beq\label{20}
\delta U^z {1\over \sqrt{g}}\dd_\mu\sqrt{g}(\rho U^\mu\delta U_z+\tau^\mu_{\ z}) =0.
\eeq

We now 
sum over equations (\ref{12}), (\ref{14}), (\ref{16}) and (\ref{20}) to obtain, after some algebra and index shifting: 
\beq\label{sum}
{1\over \sqrt{g}}  \delta U^\nu\dd_\mu[\sqrt{g}( \rho U^\mu \delta U_\nu +\tau^\mu_{\ \nu})]= -\rho \delta U^r(\delta U_0\dd_r\bar U^0 +\delta U_\phi\dd_r\bar U^\phi).
\eeq
The final step is to use $\delta ( U^\mu  U_\mu) = 0$,  which gives
$$
\delta U_0 = -{\bar U^\phi \over \bar U^0} \delta U_\phi.  
$$
Using this in (\ref{sum}), averaging $\delta U^r \delta U_\phi$ to form $\W$ and collecting terms, we obtain 
\beq\label{fluc}
{1\over \sqrt{g}}\delta U^\nu\dd_\mu[\sqrt{g}\, \rho U^\mu \delta U_\nu + \tau^\mu_{\ \nu}  )  ]= -\rho \W \bar U^0\Omega' ,
\eeq
where 
\beq
\Omega' = \dd_r(\bar U^\phi/\bar U^0)
\eeq
is just the relativistic analogue of the shear gradient (e.g. Page and Thorne 1974).   

Equation (\ref{fluc}) is a relationship for the rate at which stress extracts energy from the shear, involving via $\W$ the first-order correlated velocities
that are residual fluctuations from circular motion.   As written, however, there appears at first to be a gross mismatch:  the left side of the equation
is smaller than the right by a factor of order the ratio of the drift velocity to the rotation velocity.    But this assumes that the length scales associated with the gradients on either side of the equation are comparable.   Because the extracted free energy is in fact locally {\em dissipated,} and dissipation is dominated by the
smallest scales, the gradient length scale on the left side provides the balance from the input of the right side by being, in effect, tiny.   The analysis of nonrelativistic
discs presented in Balbus \& Hawley (1998) shows that when explicit dissipation terms are included in the energy fluctuation equation from the start, the
balance struck is between the right side of equation (\ref{fluc}) and explicit viscous (or resistive) dissipation.    These energy loss terms are unimportant for large scale
transport (and can be ignored for this purpose), but they represent the ``thermal processor'' between the extracted large-scale mechanical free energy and the
disc's radiative energy losses. 
As we have already noted, the thermal timescale over which this occurs is assumed to be rapid compared with the evolutionary time scale of the disc.   This implies that the height-integrated, volume specific source term on the right side of (\ref{fluc}) satisfies
\beq\label{strad}
-\Sigma \W \bar U^0\Omega'  = -\Sigma \W \bar U^\phi (\ln \Omega)'=2{\cal F},
\eeq
i.e. the total energy extracted over the local disk thickness is equal to the energy radiated through the upper and lower surfaces.

\subsection{Large scale evolution}

Henceforth, we
drop the bars on the $\bar U$ 4-velocities, and take these non-$\delta$ quantities to be understood as time-averaged means.  
If we integrate over height, assume axisymmetry, and ignore the pressure contributions, the equation of angular momentum conservation (\ref{am0}) now expands to:
\beq\label{25b}
0= U^0U_\phi\dd_t\Sigma +{1\over\sqrt{g}}\dd_r \left[{\sqrt{g}}\Sigma\left(U^rU_\phi +\W\right)\right] +2U_\phi {\cal F}.
\eeq
In the first term of (\ref{25b}), we have assumed that $U^0$ and $U_\phi$ are prescribed functions of $r$ only.  The final term is obtained by integrating $\dd_z \tau^z_{\ \phi}$ over height, which is now the {\em angular momentum} radiated from each side of the disc (Page \& Thorne 1974).  
Using equation (\ref{partcon}) for $U^0\dd_t\Sigma$ and simplifying, we obtain:
\beq\label{mdot}
U'_\phi \sqrt{g}\Sigma U^r +\dd_r\left( \sqrt{g} \Sigma \W\right) +2\sqrt{g}{\cal F}{U_\phi}=0,
\eeq
where $U'_{\phi}=dU_\phi/dr$.  Using now (\ref{mdot}) back in equation (\ref{partcon}), we find:
\beq\label{thex}
{\dd\Sigma\over \dd t} =  {1\over \sqrt{g}U^0}{\dd\ \over \dd r}{1\over U'_\phi}  \left[  {\dd\ \over \dd r}\left(\sqrt{g}\Sigma \W\right)+ 2\sqrt{g}{\cal F}{U_\phi} \right].
\eeq
The final step is to use equation (\ref{10b}) for ${\cal F}$.   With 
\beq\label{yy}
Y\equiv \sqrt{g}\Sigma \W,
\eeq
this brings us to our governing equation:
\beq\label{fund}
{\dd Y\over \dd t} =  {\W\over U^0}{\dd\ \over \dd r}{1\over U'_\phi} \left[   {\dd Y \over \dd r}- U_\phi U^\phi (\ln\Omega)' Y \right].
\eeq
This is the equation we have been seeking.  The first term on the right in square brackets is a straight translation from the Newtonian
equation, while the second term is a relativistic correction stemming from the photon angular momentum.    

A final point.  
We have been assuming that $\W$ is a specified function of $r$.   If, however, $\W$ has functional dependence upon
$\Sigma$, then $\W$ would be implicitly time-dependent.  In that case, equation (\ref{fund}) should be modified to: 
\beq\label{fund2}
{\dd (Y/\W)\over \dd t} =  {1\over U^0}{\dd\ \over \dd r}{1\over U'_\phi} \left[   {\dd Y \over \dd r}- U_\phi U^\phi (\ln\Omega)' Y \right],
\eeq
a form that holds more generally.

\section {Solution of the evolutionary equation}
\subsection{Preliminaries}

Let us introduce a more compact formulation.  Define $Q$ by
\beq
{dQ\over dr} = - U_\phi U^\phi (\ln\Omega)'.
\eeq
Equation (\ref{fund}) becomes
\beq\label{27q}
{\dd (Ye^Q)\over \dd t} =  {e^Q \W\over U^0}{\dd\ \over \dd r}{e^{-Q}\over U'_\phi} \left[   {\dd (Ye^Q) \over \dd r}\right].
\eeq
Next, with
\beq\label{33b}
 dH\equiv e^QU'_\phi dr, \quad \zeta =Ye^Q,
\eeq
our governing equation takes the form of a pure diffusion equation
\beq\label{pure}
{\dd \zeta \over \dd t} =  {e^{2Q} \W U'_\phi \over U^0}\ {\dd^2\zeta\over \dd H^2}.
\eeq
This has a steady-state solution of $\zeta\propto H$ .   Equations (\ref{mdot}) and (\ref{strad}) together imply
\beq
{d\zeta\over dH} =-{\sqrt g\Sigma U^r}\equiv {\dot m\over 2\pi}, \quad {\rm or}\quad \zeta = {\dot m H\over 2\pi},
\eeq
where $\dot m$ is the time-steady accretion rate and $H$ contains an additive constant boundary condition embodying the vanishing stess
location (conventionally the ISCO radius).   This is, in fact, the Novikov \& Thorne (1973) solution in its entirety!   
This reader may wish to verify this for the relatively simple Schwarzschild limit with $\Omega^2=r_g/r^3$ and
$$
  U^0=e^{-Q}=(1-3r_g/r)^{-1/2}, \quad U'_\phi =  {\Omega\over 2} {r-6r_g\over (1-3r_g/r)^{3/2}}.
$$

\subsection {Modal solution}
\subsubsection{Global WKB}

We seek time-dependent solutions of the form $e^{st}$.   Then equation (\ref{pure}) becomes
\beq\label{U2}
{d^2\zeta\over dH^2} = {s e^{-2Q} U^0\over \W U'_\phi }\zeta.
\eeq
When $\W$ is small (a not unphysical choice) or $s$ sufficiently large, equation (\ref{U2}) has the formal (unnormalised) WKB solution
(Bender \& Orszag 1978):
\beq\label{wkb1}
\zeta =  \left(e^{2Q} \W U'_\phi \over U^0\right)^{1/4}\exp\left[ \pm \int \left( s e^{-2Q} U^0\over \W U'_\phi\right)^{1/2}dH\right].
\eeq
When $s/U'_\phi<0$ we should of course interpret this in terms of trigonometric functions.   Returning to $r$ in preference to $H$, and $Y$ in preference to
$\zeta$, we obtain
\beq\label{wkb2}
Y = e^{-Q/2}  \left(\W U'_\phi \over U^0\right)^{1/4}\exp\left[ \pm \int^r  \left( s U^0U'_\phi \over \W \right)^{1/2}dr \right].
\eeq

Consider first an unstable mode, $s>0$.  At the ISCO location $r=r_I$, our WKB solution formally breaks down, but as we shall
see, it is still valid rather close to it.   Let $x=r-r_I$, so that positive and negative $x$ define regions of stable and unstable circular
orbits, with $U'_\phi >0$ and $U'_\phi<0$ respectively.   
On physical grounds we certainly should not expect much of a disc-like structure to prevail for $x<0$, but it is of interest to see how the equation discovers this on its own.  

For $x>0$, we have $U'_\phi >0$ and the solution that is well-behaved as $x\rightarrow \infty$ takes the form 
\beq\label{wkb3}
Y =  \left(\W U'_\phi \over e^{2Q}U^0\right)^{1/4}\exp\left[ - \int^r_{r_I}  \left( s U^0U'_\phi \over \W\right)^{1/2}dr \right].
\eeq
(We have chosen for later convenience a lower limit of integration to be $r_I$.    For a convergent integral, this simply amounts to setting the normalisation factor.)    In the WKB limit, this is a sharply cut-off function in the bulk of the disc $x>0$.     For $x<0$, $U'_\phi<0$ and
we may write down a formal solution:
\beq\label{wkb4}
 Y =  A \left(\W U'_\phi \over e^{2Q}U^0\right)^{1/4}\,\,\, \sin \left[ \int^r_{r_I}  \left( -s U^0U'_\phi \over \W\right)^{1/2}dr  \,+\, \Phi\right].
\eeq
Here, the amplitude $A$ and phase $\Phi$ are determined by the requirement that the $x<0$ solution join smoothly onto the exponentially
cut-off solution for $x>0$.   This is already enough to see that unstable modes have significant amplitudes only inside of the ISCO,
a physically very sensible result.    

For stable ($s<0$) solutions, it is clear from our general solution (\ref{wkb2}) that the exponential cut-off behaviour now occurs inside the ISCO, $x<0$, while for $x>0$, the bulk of the disk hosts a spectrum of spatially-oscillatory, temporally-decaying modes.   

\subsubsection{Local ISCO structure for nanvanishing ${\W}$}

The ISCO  is an apparent singularlity of our equation, which can, however, be treated rigorously.   
In the local vicinity of the ISCO $r=r_I$ equation (\ref{27q}) may be written
\beq
sY = \left(W^r_{\ \phi}\over U^0U''_\phi\right)_{\!I}{d\over dx}\left( {1\over x}{dY\over dx}\right),
\eeq
where $x=r-r_I$,  $U'_\phi(r)=U''_\phi (r_I)x$, and the notation $()_I$ means that $W^r_{\ \phi}$, $U^0$, and $U''_\phi$ are all evaluated at the ISCO $r=r_I$.   
Note that the $Q$ term is subdominant and has actually disappeared from the local ISCO-centred equation.
(It will reappear as part of a locally determined normalisation factor.)
We shall first assume that $\W(r_I)$ does not vanish, with the ultimate intent of showing
the opposite: on physical grounds, it must vanish in a thin disc.    With finite $\W(r_I)$, the (unnormalised) solution to this equation for $s>0$ is
\beq\label{ai}
Y={\rm Ai}'(k x), \quad k\equiv \left(sU^0 U''_\phi\over W^r_{\ \phi}\right)^{1/3}_{\!I},
\eeq
where ${\rm Ai}'$ is the derivative of the Airy function.    As in our WKB solution (\ref{wkb3}), positive values of the argument
correspond to exponentially cut-off behaviour (the solution not chosen, Bi$'$, rises exponentially), whereas negative values
correspond to oscillatory behaviour.   (See figure [1].)  The ``dispersion relation'' we have found, the $k$-definition of equation (\ref{ai}), may be written
\beq\label{disp}
s= \left( W^r_{\ \phi}\over U^0U''_\phi\right)_I k^3,
\eeq
and exhibits violent instabilites on the smallest scales.   This is a compelling reason to seek physically viable solutions with the ISCO boundary condition $W^r_{\ \phi}=0$.   

Before we do, however, we note a point of some mathematical consequence.   The WKB solution (\ref{wkb1}) depends upon large $|sU^0U'_\phi /W^r_{\ \phi}|$ for its validity, whereas the local solution merely requires $x\ll r_I$.  {\it These are not mutually exclusive restrictions.}  There is no reason why they both cannot be valid in an overlapping domain.   In this shared asymptotic regime, the two solutions must take one and the same form.
To verify this is indeed so, note that the large argument expansion of the Ai$'$ function is (up to an overall normalisation):
\beq\label{41a}
{\rm Ai}'(kx)\rightarrow x^{1/4} \exp\left[-{2\over 3}(kx)^{3/2}\right],
\eeq
which is exactly the same form as equation (\ref{wkb3}) in the limit $r\rightarrow r_I$, $U'_\phi\rightarrow U''_\phi x$ (once again up to an overall normalisation):
$$
\left( W^r_{\ \phi}\ U'_\phi\over e^{2Q} U^0\right)^{1/4} \exp \left[ -\int_0^x\left( sU^0U'_\phi \over W^r_{\ \phi } \right)^{1/2}\, dx\right]\rightarrow
$$
\beq\label{42b}
{\rm constant}\ \times\ x^{1/4}\exp\left[-{2\over 3}(k x)^{3/2}\right].
\eeq
Equations (\ref{41a}) and (\ref{42b}) have exactly the same functional form, as was sought.

\subsubsection {A uniformly valid solution}

The agreement between the two solutions in an overlapping asymptotic zone suggests the possibility that there may be a single analytic formula that is valid everywhere.  Such a solution is known to exist for a certain class of ``one-turning-point problems,'' in quantum mechanical solutions of the Schr\"odinger equation (Bender \& Orszag 1978).   Rather than derive this function, it is simplest just to write it down, and then verify that it reduces to each of our asymptotic forms in the appropriate limits.  

Define $X$ by
\beq\label{XX}
X= \left[ {3\over 2} \int_{r_I}^r \left( sU^0U'_\phi\over \W\right)^{1/2}dr\right]^{2/3}.
\eeq
Then, our (unnormalised) uniformly valid solution is 
\beq\label{uni}
Y = e^{-Q/2}\left( \W U'_\phi\over XU^0\right)^{1/4} {\rm Ai}'(X).
\eeq

To verify this, we assume first that $s>0$.  In the limit $X\gg 1$, Ai$'(X)$ has the asymptotic form 
$$
{X^{1/4} \over 2\sqrt{\pi}} \exp \left(-{2\over 3} X^{3/2}\right) =  \qquad\qquad\qquad\qquad\qquad \ \ \ \ \ \ \ \ \ \ \
$$
\beq
\qquad\qquad \qquad \ \ \ \ \ \ \ \ {X^{1/4}\over2\sqrt{\pi}}\exp \left[ -\int_{r_I}^r \left( sU^0U'_\phi \over \W  \right)^{1/2}\, dr\right],
\eeq
so the $X^{1/4}$ factors cancel in (\ref{uni}), and we are led directly to equation (\ref{wkb3}) for $Y$.   Next, when $r\rightarrow r_I$ and $U'_\phi>0$, we expand $U'_\phi=xU''_\phi$,
and $X$ becomes
$$
X=\left[ {3\over 2} \left( sU^0U''_\phi\over \W\right)_{\!I}^{1/2}\int_0^x x^{1/2}\,dx\right]^{2/3} = kx,
$$
and $(\ref{uni})$ then reduces to equation (\ref{ai}):  $Y\sim {\rm Ai}'(kx)$, since $U'_\phi/X$ is locally equal to the constant $U''_\phi/k$.  Finally, when $x<0$ away from the ISCO, then $U'_\phi<0$.  Multiply $U'_\phi$ by unity, written as $-e^{i\pi}$.   Then,
$$
X=  e^{i\pi/3}\left[ {3\over 2} \int_{r_I}^r \left(-{  sU^0U'_\phi\over \W}\right)^{1/2}dr\right]^{2/3}.
$$
Switching the limits $r$ and $r_I$, this is the same as
$$
X = -\left[ {3\over 2} \int_r^{r_I} \left( - sU^0U'_\phi\over \W\right)^{1/2}dr\right]^{2/3}<0,
$$
i.e, $X$ is a purely real negative quantity, despite all of the complex-valued exponents and nested fractional powers.   Then, use the standard large negative argument for Ai$'(X)$ (Bender \& Orszag 1978):
\beq
{\rm Ai}'(X)\rightarrow {(-X)^{1/4}\over \sqrt{\pi}}\sin\left[ {2\over 3}(-X)^{3/2} + {\pi\over 4}\right].
\eeq
It is easy to see that with equation (\ref{uni}), by adjusting $A$ and $\Phi$ the above leads to a precise match with equation (\ref{wkb4}).
Equation (\ref{uni}) is therefore a uniformly valid solution to equation (\ref{fund}).   Figure (1) shows an explicit solution for Schwarzschild geometry with $\W\propto r^{1/2}$, chosen for ease of analytics.     (Recall that $\W$ correlates angular momentum and radial velocity fluctuations, so there is some sense for it to increase slowly with $r$.)  In this case, the needed integral over $(U^0U'_\phi/\W)^{1/2}$ is
(see the end of \S 3.1): 
$$
\int {\sqrt{r'-6}\over {r'-3}}dr' =2\sqrt{r'-6} - 2\sqrt{3}\tan^{-1}\left(r'-6\over 3\right)^{1/2}
$$
where $r'$ is $r/r_g$.

We conclude with a final formula for the disc surface density $\Sigma(r,t)$:
\beq
\Sigma = e^{-Q/2}\, \left( U'_\phi\over g^2(\W)^3  XU^0\right)^{1/4} {\rm Ai}'(X)\, \exp(st).
\eeq
with $X$ given by (\ref{XX}).

\begin{figure}
	\centering
           \includegraphics[width=6cm,clip=true, trim=0cm 0cm 0cm 0cm]{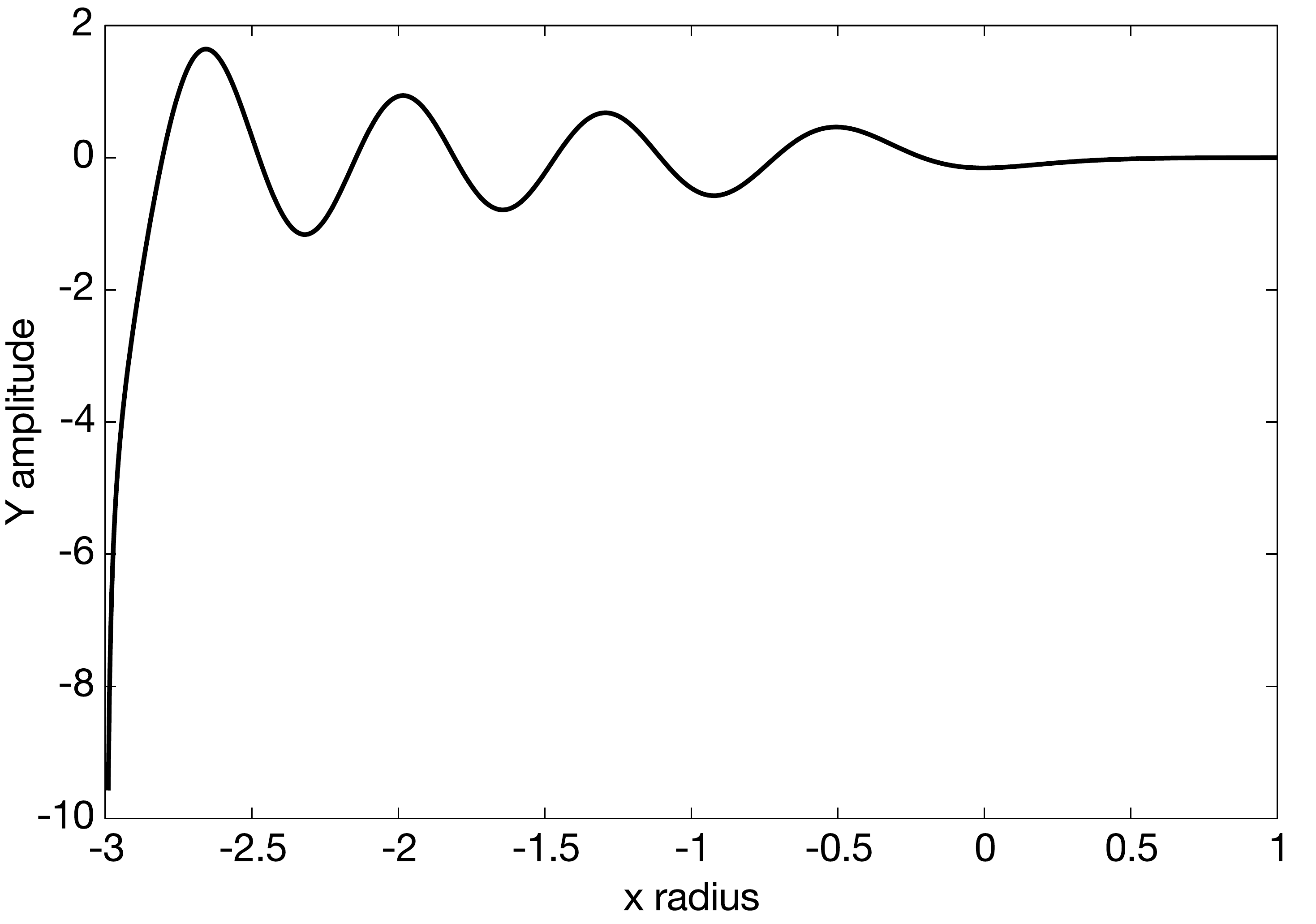} 				
	\caption{Plot of the function $Y$, an unstable mode near the ISCO, located at $x=0$ in the figure.   The mode shown corresponds to $\W\propto r^{1/2}$, chosen for ease of calculation.  The spatial response is confined almost entirely to the region $x<0$, where the angular momentum $U_\phi$ increases inward.   Note the singular behaviour near the innermost photon orbit at $x=-3$.}
\label{fig1}
\end{figure}

\subsection{Modal solutions for vanishing $\W(r_I)$}
\subsubsection{ Exterior region, $r>r_I$}

If there is any finite stress at the the ISCO, then there are extremely unstable modes present on small scales. 
This is a compelling argument in favour of the customary boundary condition of setting $\W=0$ for $x\le0$.    Let us see
how this removes the unstable behaviour.   

We shall assume that $\W(r_I)$ vanishes.  In equilibrium, $\zeta\propto H$.  From the definitions of $H$ and $\zeta$ in (\ref{33b}), it
follows that $\W \propto x^2$ .
The question then is what are the solutions of (\ref{fund}) near the ISCO with this behaviour for $\W$?

Set the local stress $\W=\W''x^2/2$.  Then, the local ISCO equation is
\beq
sY = \left({\W}''\over 2U^0U''_\phi\right)_{\!I}x^2{d\over dx}\left( {1\over x}{dY\over dx}\right),
\eeq
or
\beq
Y''-{Y'\over x} + \beta {Y\over x}=0, \qquad \beta=- \left(2sU^0U''_\phi\over \W''\right)_{\!I}.
\eeq
By equation (\ref{yy}), $Y$ itself must now vanish at $x=0$.  If $s>0$, then there are two formal solutions to this equation in the region $x\ge 0$:
\beq
xI_2(2\sqrt{|\beta| x}) , \quad xK_2(2\sqrt{|\beta |x}).
\eeq
But the $I_2$ solution is not well-behaved for large $x$, and the $K_2$ solution does not vanish at $x=0$, so in fact there are
no solutions compatible with the boundary conditions.  In other words, {\it there are no unstable $s>0$ solutions for $x\ge 0$.}  

Consider next $s<0$.    Then, the solution satisfying the vanishing $Y$ boundary condition at $x=0$ is
\beq\label{J2}
Y=xJ_2(2\sqrt{\beta x}),
\eeq
where $J_2$ is the Bessel function of order $2$.  
The corresponding solution with the $Y_2$ Bessel function does not vanish at $x=0$.
Hence, there is a well-determined stable set of modal solutions with vanishing $\W$ at the ISCO.   

Once again, there is an overlap zone near the ISCO in which the WKB solution is valid together with the small $x$ local solution.
The large argument
expansion of (\ref{J2}) is
\beq\label{largj}
Y\rightarrow -\left({2x^3\over \pi^2\beta}\right)^{1/4} \sin \left(2\sqrt{\beta x}+{\pi\over 4}\right).
\eeq
The WKB solution follows from (\ref{wkb4}), now {\it outside} the ISCO:
\beq\label{sinus2}
Y= A\left(- U'_\phi \W\over e^{2Q} s U^0\right)^{1/4}\! \sin\left[\int^{r}_{r_I} \left(- sU^0U'_\phi \over \W \right)^{1/2}\! dr \!+\! \Phi\right],
\eeq
where $A$ and $\Phi$ are once again arbitrary.      In the limit $r\rightarrow r_I$, we require $U'_\phi\rightarrow U''_\phi x$  and $\W\rightarrow\W''x^2/2$.  The integral in (\ref{sinus2}) is then precisely $2\sqrt{\beta x}$.   With the proper choice of $A$ and $\Phi$, there is a complete agreement of functional form between (\ref{sinus2}) and (\ref{largj}).   

Finally, there is once again a simple, uniformly valid solution.  With $X$ now defined by 
\beq
X = \int^{r}_{r_I} \left(U^0U'_\phi \over \W \right)^{1/2}\! dr,
\eeq
the desired $r>r_I$ solution is
\beq\label{YYY}
Y = e^{-Q/2}\left({U_\phi' X^2\W\over U^0}\right)^{1/4}J_2(\sqrt{-s}X). 
\eeq
To verify this, simply expand the above:  first for large $\sqrt{-s}X$ (recovering [\ref{sinus2}]), then for small $x$ (recovering [\ref{J2}]), and then simultaneously for large $X$ and small $x$ (recovering [\ref{largj}]).   It is readily seen that the function (\ref{YYY}) reduces to all proper asymptotic forms.   

\subsubsection{Interior region, $r<r_I$}

For $r<r_I$, there are no {\it stable} solutions that are well-behaved.   The unstable $s>0$, but spatially well-behaved, interior solution is now easy to construct, since it has precisely the same mathematical form as the exterior solution.   Moreover, vanishing at $x=0$ together with its first derivative, this solution seems to join smoothly onto the stable exterior solution.   The smoothness is maintained even though
the growth rate is different on either side of $x=0$!   How does
it make sense to have a global ``mode'' with two different growth rates, one with positive $s$, the other with negative $s$, in different regions?    

Of course a single mode cannot have different growth rates in different disk regions.   What we have been discussing is in reality a superposition of two modes.
This points to the resolution of our problem.   The location $x=0$ is a branching singularity of the governing equation, and lacks a unique prescription for traversing it.   It is an ``improper node.''   All modal solutions have vanishing $Y$ and $dY/dx$ at $x=0$.  In particular, a smooth solution to our problem is
that equation (\ref{YYY}) holds for $x>0$, and $Y=0$ for $x<0$, a stable mode that lives entirely in the exterior bulk of the disk.
Similar considerations hold for its ``dual,'' unstable solution,  in this case with vanishing $Y$ for $x\ge 0$.   Thus the answer to the question posed at the end of the previous paragraph is that the peculiar global solution described is not, in fact, one mode,
but a superposition of two.   What is perhaps unusual is that each mode vanishes identically where the other is present!    
The stable exterior mode is unique, and for astrophysical purposes, {\em the} mode of interest.     

\subsection{Green's function solution}

With $Y$ given by equation (\ref{YYY}), we may construct more general solutions by superposing modes.  Formally, we may write
\beq\label{G1}
Y(r,t) = \int A(s)\, Y(\sqrt{s})e^{-st}\, ds
\eeq
where $A(s)$ is whatever appropriate function we choose.   (For convenience, we have replaced $s$ by $-s$ in this role as a dummy variable, and the explicit $s$ dependence is exhibited in $Y$.)    Consider next the integral (Gradshteyn \& Ryzhik 2014):
$$
\int^\infty_0 J_p(\sqrt{s}X)J_p(\sqrt{s}X_0)  e^{-st}\, ds = 
$$
\beq\label{G2}
{1\over t}\exp\left(-X^2-X_0^2\over 4t\right)I_p\left(XX_0\over 2t\right),
\eeq
where $J_p$ is the Bessel function of order $p$ and $I_p$ the corresponding modified Bessel function.  
In the limit $t\rightarrow 0$, this integral represents a delta function type of concentration at $X=X_0$ which then spreads as $t$ increases.  This behaviour, together with $XX_0$ symmetry, is what we seek for a Green's function response, initially concentrated
at $X=X_0$.   We are assuming that our global WKB solution holds over all $s$ in the integral, an assumption that must break down at $s=0$.   
But in the limit
of small $\W$, this will affect only the detailed behaviour at very late times; the small $s$ contribution to the integral is otherwise negligible.

Combining the results of equations (\ref{YYY}), (\ref{G1}), and (\ref{G2}) allows us to write down the (unnormalised) Green's function
solution to our equation.  With $X=X(x)$ and $X_0=X(x_0),$
$$
G(x, t;x_0) = \left({U_\phi' X^2\W\over e^{2Q} U^0}\right)_{x=x_0}^{1/4} \left({U_\phi' X^2\W\over e^{2Q} U^0}\right)^{1/4}  \times
$$
\beq\label{Gf}
{1\over t}  \exp -\left[(X-X_0)^2\over 4t\right]e^{-XX_0/2t}\, I_2\left(XX_0\over 2t\right)
\eeq
At early times $t\rightarrow 0$, the asymptotic behaviour of the terms on the final line of (\ref{Gf}) simplifes to:
\beq
\rightarrow
{1\over (\pi t XX_0)^{1/2} } \exp -\left[(X-X_0)^2\over 4t\right].
\eeq
This takes on the classic form for the diffusion of an initially very localised concentration.
The {\it local-frame} surface emissivity is then given directly by (\ref{10b}):
\beq
{\cal F} = -{1\over 2\sqrt{g}} G(x,t;x_0) U^0\Omega'.
\eeq

\section{Discussion}

In this paper, we have derived a form of the thin disc diffusion equation that is valid for general relativisitc spacetimes.   We assume only that the metric tensor is axisymmetric, so that 
our equation is suitable for both the Schwarzschild and Kerr geometries.   Remarkably, the metric itself enters into
the calculation only in the form of a determinant, which is then absorbed as a multiplicative factor of our surface density variable.   It then disappears entirely from the calculation.

The physics of an evolving thin relativistic disc compels unstable modes trapped inside $r=r_I$ to rapidly destroy their host
in this ``Rayleigh-unstable'' zone.    Quasi-stable equilibrium circular orbits are mathematical fantasies here:  without a retaining potential, the orbits simply plunge.    The exterior modal solutions, by contrast, are always stable.   On the other hand, by imposing the boundary condition of a vansihing stress tensor at $x=0$  ($\W \sim x^2$), stable modes exist {\it exclusively} in the stable region, vanishing together with their first derivatives as $x\rightarrow 0$ from positive values.   Modes with finite $\W$ at $x=0$ must penetrate, at least exponentially, into the plunging region, and in thin disc models would not be supported.

By using a combination of WKB techniques, local analysis and matched asymptotic expansions, it is possible to solve very generally the disc diffusion equation in terms of quadratures.   This is the scope of the current paper.   Using these methods, we have been able to calculate the Green's function solution, which is
expected to be valid up until very late times when a quasi-equilibrium is reached.    These findings may be useful
as numerical diagnostics, but only if the thin disc condition is well-satisfied, a limit that has yet to be convincingly simulated.      The most interesting astrophysical application of this work is likely to be to black hole transients.    These include dramatic state changes in which the inner regions of the disc are thought to
disappear and then reform, and tidal disruption events in which a disc forms and subesquently accretes from the debris of a mangled star.   In principle, these events may be modelled by the one-dimensional evolution equation (\ref{fund}) with appropriate boundary conditions, and the time-dependent surface emission calculated in the observer's reference frame.   These interesting possibilities will be the subject of future investigations.



\section*{Acknowledgements}

It is a pleasure to acknowledge important discussions with W. Potter during the early formative stages of this work, and very constructive
comments from C. McKee, M. Rees and C. Gammie on an earlier manuscipt.    I am grateful for support from the Royal Society in the form of a Wolfson Research Merit Award, and from STFC (grant number ST/N000919/1).



\label{lastpage}

\end{document}